\documentclass[lettersize,journal]{IEEEtran}
\IEEEoverridecommandlockouts
\usepackage{cite}
\usepackage{amsmath,amssymb,amsfonts}
\usepackage{algorithm}
\usepackage{algpseudocode}
\usepackage{graphics}
\usepackage{epsfig}
\usepackage{epstopdf}
\usepackage{textcomp}
\usepackage{xcolor}
\usepackage{multirow}

\def\BibTeX{{\rm B\kern-.05em{\sc i\kern-.025em b}\kern-.08em
    T\kern-.1667em\lower.7ex\hbox{E}\kern-.125emX}}
\begin{document}
\bstctlcite{BSTcontrol}
\title{Patterned Beam Training:  A Novel Low-Complexity and Low-Overhead Scheme for ELAA} 


\author{  
	Hongkang~Yu,
  Yuan~Si,
  Shujuan~Zhang,
	and~Yijian~Chen
}	

\maketitle

\begin{abstract}
  Extremely large antenna arrays (ELAAs) can provide higher spectral efficiency. However, the use of narrower beams for data transmission significantly increases the overhead associated with beam training. In this letter, we propose a novel patterned beam training (PBT) scheme characterized by its low overhead and complexity. This scheme requires only a single linear operation by both the base station and the user equipment to determine the optimal beam, reducing the training overhead to half or even less compared to traditional exhaustive search methods. Furthermore, We discuss the pattern design principles in detail and provide specific forms. Simulation results demonstrate that the proposed scheme outperforms the compared  methods in terms of beam alignment accuracy and achieves a balance between signal-to-noise ratio (SNR) conditions and training overhead, making it a promising alternative.
\end{abstract}

\section{Introduction}
\IEEEPARstart{F}{uture} mobile communication systems will operate at higher frequency bands to provide greater bandwidth and higher data rates \cite{ref1}. To combat severe pathloss, the base station (BS) will be equipped with an extremely large antenna array (ELAA), utilizing beamforming techniques to ensure coverage \cite{ref2}. The increased array aperture leads to narrower beams, making beam alignment between the BS and the user equipment (UE) crucial for system performance.

Typically, the BS performs beam training to achieve beam alignment \cite{ref3}. One common approach is to exhaustively sweep through all candidate beams and determine the optimal one according to UE feedback. Unfortunately, the sharp increase in training overhead has limited its application in ELAA scenarios.

To address this issue, hierarchical beam training scheme has been proposed \cite{ref4, ref5}. This scheme involves transmitting training beams with varying widths across multiple stages. Nevertheless, due to the training beam cannot be shared by all UEs, this approach exhibits limited efficiency in multi-user scenarios.
 Moreover, \cite{ref6, ref6_1} leverage compressive sensing (CS) theory to recover channel state information (CSI), and the optimal beam can be identified naturally. However, this method needs iterative operations for sparse signal recovery, leading to high complexity.
 Additionally, a multi-armed beam training strategy is adopted in \cite{ref7, ref8, ref9}, where each training beam incorporates multiple main lobes. Through cross-validation among these training beams, the optimal beam can be rapidly obtained. Our previous research \cite{ref9} has analyzed the impact of training beam design on performance, but there is still room for optimization.

 In this letter, we propose a novel scheme called patterned beam training (PBT). Based on the probe pattern and combining pattern designed in the scheme, both BS and UE need to perform only a single linear processing over training pilots to detect the optimal beam. The proposed scheme offers the advantages of low overhead and complexity, while effectively supporting multi-user scenarios. To the best of our knowledge, it can serve as a promising alternative approach. 
 
 Furthermore, we conduct a detailed analysis of the pattern design principles and propose multiple specific forms of patterns, where exhaustive and multi-armed beam training schemes can be regarded as special cases of PBT. 
 Simulation results indicate that the PBT can significantly reduce training overhead and achieve a balance between overhead and signal-to-noise ratio (SNR) conditions compared to the exhaustive method. Additionally, the proposed scheme exhibits superior beam alignment accuracy compared to other approaches.

\section{System Model}
Consider a multi-input single-output (MISO) system, where the BS is equipped with a uniform linear array (ULA) with $N$ antenna elements. This letter focuses on the beam training process between the BS and a single-antenna UE. Given that the proposed scheme could be shared by all UEs in the cell, it can be easily extended to multi-user scenarios. Assuming that the UE remains stationary or moves at a low speed, the channel is considered fixed during the beam training process. We adopt a geometry-based model to represent the sparsity of high-frequency channels, which can be expressed as
\begin{equation}
	{\mathbf{h}} = {\alpha _0}{\mathbf{a}}\left( {{\theta _0}} \right) + \sum\limits_{p = 1}^P {{\alpha _p}{\mathbf{a}}\left( {{\theta _p}} \right)},
\end{equation}
where ${\mathbf{a}}\left( \theta  \right) = {\left[ {1,{e^{ - j\pi \theta }}, \ldots ,{e^{ - j\pi \left( {N - 1} \right)\theta }}} \right]^{\text{T}}}/\sqrt N ,\theta  \in \left[ { - 1,1} \right]$ denotes the steering vector of the ULA, $\alpha_0 [\alpha_p]$ and $\theta_0[\theta_p]$ represent the gain and the angle of the line-of-sight (LoS) [$p$th non-LoS (NLoS)] path, respectively. Here, $P$ is the total number of NLoS paths.

In high-frequency channels, ${\alpha _0} \gg {\alpha _p}$, and the goal of beam training is to select the optimal beam from candidate beams for data transmission, maximizing the beam gain in $\theta_0$. We suppose that the set of candidate beams consists of $N$ discrete Fourier transform (DFT) beams. The beamforming weight of the $n$-th ($n=1,2,\dots,N$) beam is denoted as $\bf d_n$, corresponding to the $n$-th column of the $N$-dimensional DFT matrix $\bf D$, where ${d_{n,m}} = {\exp({ - j2\pi \left( {n - 1} \right)\left( {m - 1} \right)/N})}/\sqrt N $.

In traditional exhaustive beam training schemes, the BS sequentially uses the aforementioned DFT beams to transmit pilot signals. The symbols received by the UE under $N$ training beams can be expressed as
\begin{equation}\label{eq2}
	{\mathbf{y}} = {{\mathbf{D}}^{\text{H}}}{\mathbf{h}} + {\mathbf{v}},
\end{equation}
where ${\mathbf{v}}\sim\mathcal{C}\mathcal{N}\left( {{\mathbf{0}},\sigma _{\text{v}}^2{\mathbf{I}}} \right)$ denotes the noise vector, and $\sigma _{\text{v}}^2$ is the normalized noise power.

To estimate the optimal beam index $n^*$, the UE compares the received signal power ${\left| {{y_n}} \right|^2}$ of each training beam and feeds the estimation result $\hat n = \arg \mathop {\max }\limits_n {\left| {{y_n}} \right|^2}$ back to the BS. Subsequently, the BS will use the corresponding beam ${{\mathbf{d}}_{\hat n}}$ to transmit data. According to the above description, the overhead of exhaustive beam training is proportional to the number of antennas. Therefore, this approach faces significant challenges in future ELAA scenarios.

\section{Proposed Patterned Beam Training Scheme}
To reduce the overhead of beam training, this section introduces the proposed PBT scheme, which offers the advantages of low overhead and low complexity. First, we describe the overall framework of PBT, followed by the principles for pattern design within the scheme. Finally, several specific forms of patterns are proposed.
\subsection{Framework of Patterned Beam Training}
In PBT, the BS transmits $M<N$ training beams. We define the \emph{\textbf{probe pattern}} ${\mathbf{B}} = \left[ {{{\mathbf{b}}_1},{{\mathbf{b}}_2}, \ldots ,{{\mathbf{b}}_M}} \right]$, which is composed of the weights $\mathbf{b}_m$ of the training beam. The received signal at the UE side can be represented as
\begin{equation}
{\mathbf{r}} = {{\mathbf{B}}^{\text{H}}}{\mathbf{h}} + {\mathbf{z}}, 
\end{equation}
where $\bf z$ is an $M$-dimensional noise vector, having the same distribution as $\bf v$. 

To select the optimal beam from $N$ candidate beams, a key operation in PBT is the linear dimensionality expansion of $\bf r$. Specifically, the BS requires the UE to process $\bf r$ using the corresponding \emph{\textbf{combining pattern}} ${\mathbf{\tilde B}} \in {\mathbb{C}^{N \times M}}$ according to the selected $\bf B$, which results in\footnote{This letter focuses on the processing of complex signals $\bf r$, requiring coherence between different pilot symbols. For certain pattern designs, processing the magnitude 
$|\bf r|$ can also identify the optimal beam. Generally, coherent processing methods offer better performance.}
\begin{equation}\label{eq3}
	{\mathbf{\tilde y}} = {\mathbf{\tilde Br}} = {\mathbf{\tilde B}}{{\mathbf{B}}^{\text{H}}}{\mathbf{h}} + {\mathbf{\tilde z}},
\end{equation}
where ${\mathbf{\tilde z}} = {\mathbf{\tilde Bz}}$ is colored noise. Similarly, by comparing the magnitude of each element in ${\mathbf{\tilde y}}$, the UE can provide feedback to the BS regarding the estimated optimal beam.

By comparing \eqref{eq2} and \eqref{eq3}, it can be implied that through sophisticated pattern design to make ${\mathbf{\tilde B}}{{\mathbf{B}}^{\text{H}}} \approx {{\mathbf{D}}^{\text{H}}}$, PBT can achieve functionality similar to the exhaustive beam training. Unfortunately, due to $M<N$, we have ${\text{rank}}( {{\mathbf{\tilde B}}{{\mathbf{B}}^{\text{H}}}} ) < N$ and ${\mathbf{\tilde B}}{{\mathbf{B}}^{\text{H}}}$ does not strictly equal $\bf D^\text{H}$. Therefore, PBT cannot achieve the same performance as exhaustive beam training. Nonetheless, pattern design has a significant impact on the accuracy of beam alignment in the proposed scheme, which will be detailed in the following subsection.

\textbf{Remark 1} (\emph{Comparison of PBT and existing schemes}): The primary innovation of the proposed approach lies in obtaining the variable ${\mathbf{\tilde y}}$ for determining the optimal beamforming solely through a single linear transformation ${\mathbf{\tilde B}}$. In contrast, for CS approaches like the orthogonal matching pursuit (OMP) algorithm, iterative processing of $\bf r$ is necessary. Meanwhile, AI-based beam training schemes require the computation of ${f_{{\text{NN}}}}\left( {\mathbf{r}} \right)$, where ${f_{{\text{NN}}}}\left( \cdot \right)$ represents a neural network. Both approaches involve nonlinear processing of the received signals. Supported by simulations, we believe that if the goal is solely to identify the optimal beam rather than to accurately estimate the channel vector $\bf h$, the proposed scheme relying on linear transformations already exhibits sufficiently good performance. 

\subsection{Pattern Design Principles}
This subsection analyzes the principles for designing probe and combining patterns from the perspective of beam gain. Consider a normalized beamforming weight $\bf w$, which induces the beam gain ${\left| {{{\mathbf{w}}^{\text{H}}}{\mathbf{a}}\left( \theta  \right)} \right|^2}$ at $\theta$. In our analysis, we focus on the beam gain at discrete grid points  $\theta = - 1 + 2i/N, i=0,1,\dots,N-1$, which can be calculated using the DFT matrix, i.e., ${\mathbf{g}} = {\left| {{{\mathbf{w}}^{\text{H}}}{\mathbf{D}}} \right|^2}$. For exhaustive beam training schemes, we have $\mathbf{g}=\mathbf{e}_i$, where $\mathbf{e}_i$ is the selection vector. This implies that the power of DFT training beams is concentrated in the main lobe direction.

In the case of PBT, according to \eqref{eq3}, after the combining operation, it can be considered to have $N$ equivalent training beams with weights ${{\mathbf{w}}_n} = {\mathbf{B}}{{\mathbf{\tilde b}}_n}$, where ${{\mathbf{\tilde b}}_n}$ is the  $n$-th column of ${{\mathbf{\tilde B}}^{\text{H}}}$. To ensure a fair comparison, we constrain ${\| {{{\mathbf{b}}_m}} \|_2} = 1$ and ${\| {{{{\mathbf{\tilde b}}}_n}} \|_2} = 1$, thereby limiting the transmission power and the noise amplification factor. Analogous to the beam gain of DFT training beams, we define the gain matrix ${\mathbf{G}} = {\left| {{\mathbf{\tilde B}}{{\mathbf{B}}^{\text{H}}}{\mathbf{D}}} \right|^2}$  of PBT and propose the following principles for pattern design.

\begin{itemize}
  \item{\emph{P1: Equal Main Lobe Gain}}
  
  We aim for the main lobe gains of each equivalent training beam in PBT to be equal, ensuring that UEs covered by different beams can select the optimal beam with equal probability, which corresponds to 
  \begin{equation}
    \min \;{\left\| {{\text{diag}}\left( {\mathbf{G}} \right) - g{\mathbf{1}}} \right\|_2},
  \end{equation}
  where $g = {\left\| {{\text{diag}}\left( {\mathbf{G}} \right)} \right\|_1}/N$.

  \item {\emph{P2: Minimize Power Leakage of Sidelobe}}
  
  Ideal training beams have low sidelobes. Therefore, it is desired that all sidelobe leakage power be minimized, and this can be represented by the sum of off-diagonal elements of matrix $\bf G$, i.e.,
  \begin{equation}
    \min \;\frac{1}{{{{\left( {N - 1} \right)}^2}}}\sum\limits_n {\sum\limits_{n' \ne n} {{{\left[ {\mathbf{G}} \right]}_{n,n'}}} }.
  \end{equation}
  \item {\emph{P3: Suppress Maximum Side Lobe}}
  
  The last principle focuses on maximum side lobe suppression, corresponding to
  \begin{equation}
    \min \;\mathop {\max }\limits_{n,n' \ne n} \;{\left[ {\mathbf{G}} \right]_{n,n'}}.
  \end{equation}
\end{itemize}

\begin{figure}[t]
	\centering
	\includegraphics[width=3.2in]{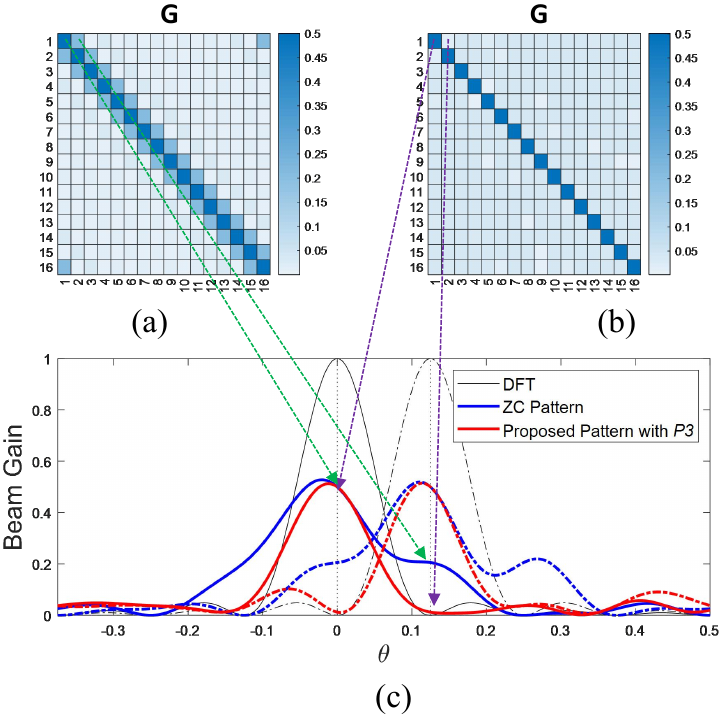}
	\caption{Relationship between gain matrix $\bf G$ and (equivalent) training beams with $N=16$, $M = 8$. (a). $\bf G$ for the ZC pattern; (b). $\bf G$ for the proposed pattern that considers \emph{P3}; (c). Comparison of gains for different training beams.}
	\label{fig1}
\end{figure}

\textbf{Remark 2} (\emph{The necessity of P3}): It's worth noting that \emph{P3} is not redundant compared to \emph{P2}. Here's an example to illustrate this. If we constrain $\bf{b}_m$ to be a constant modulus sequence and require ${\mathbf{\tilde B}} = \sqrt {N/M} {{\mathbf{D}}^{\text{H}}}{\mathbf{B}}$,\footnote{The coefficient is used to ensure ${\| {{{{\mathbf{\tilde b}}}_n}} \|_2} = 1$.} it can be verified that using Zadoff–Chu (ZC) sequences as the probe pattern can simultaneously optimize \emph{P1} and \emph{P2}. The ZC sequences is given as ${b_{1,n}} = {\exp({{{j\pi {{\left( {n - 1} \right)}^2}}}/{N}})}$, and  $\mathbf{b}_{m+1}$ is obtained by cyclically shifting $\mathbf{b}_{m}$. Unfortunately, as shown in Fig. \ref{fig1}, the main lobe of the equivalent training beams corresponding to the ZC pattern widens compared to the DFT beams, leading to a significant decrease in beam alignment accuracy. In contrast, considering \emph{P3} additionally ensures that the beam width of the proposed pattern is closer to that of the DFT beams.

In addition, the above example also illustrates the challenge in finding the optimal pattern that simultaneously minimizes the three proposed objective functions. Therefore, in the next subsection, we will design patterns based on a heuristic method and compare the performance of different patterns through Monte Carlo simulations.

\subsection{Specific Pattern Designs}
This subsection introduces the specific designs of five patterns. The first two correspond existing schemes, used to demonstrate that PBT is a general framework. The subsequent three are our proposed designs with novelty.

\emph{\textbf{Pattern 1: Exhaustive beam training scheme}}
  
  It is evident that the exhaustive method can be considered as a special case of PBT, corresponding to ${\mathbf{B}} = {\mathbf{D}}$ and ${\mathbf{\tilde B}} = {\mathbf{I}}$.

\emph {\textbf{Pattern 2: Traditional multi-armed beam training scheme}}

In the traditional multi-armed training scheme, each training beam has multiple main lobes, with its weight being the superposition of $K$ DFT vectors, i.e., ${{\mathbf{b}}_m} = \sum\nolimits_{n \in {\mathcal{C}_m}} {{{\mathbf{d}}_n}} /\sqrt K $, where ${\mathcal{C}_m}$ denotes the set of indices of $K$ selected DFT vectors for the $m$-th training beam. To select the optimal DFT beam, the UE performs a voting mechanism, with details available in \cite{ref9}.

From the perspective of PBT scheme, the probe pattern is derived as ${\mathbf{B}} = {\mathbf{DC}}$, where $\bf C$ is the hash codebook described in \cite{ref9}, which can be expressed as

\begin{equation}
  {c_{n,m}} = \left\{ \begin{gathered}
    1/\sqrt K ,\;n \in {\mathcal{C}_m}, \hfill \\
    0,\;{\text{otherwise}}. \hfill \\ 
  \end{gathered}  \right.
\end{equation}
Moreover, the corresponding combining pattern for the voting mechanism is given by ${\mathbf{\tilde B}} = \sqrt {N/M} {\mathbf{C}}$.

\emph {\textbf{Pattern 3: Proposed multi-armed beam training scheme with improvement}}

According to the pattern design principle \emph{P3}, an improved multi-armed training scheme can be easily identified by adding a random phase shift factor before $c_{n,m}$, i.e., ${\tilde c_{n,m}} = {e^{j{\phi _{n,m}}}}{c_{n,m}}$, where $\phi_{n,m}$ follows a uniform distribution over $[0,2\pi]$. The probe pattern and combining pattern remain ${\mathbf{B}} = {\mathbf{D\tilde C}}$  and ${\mathbf{\tilde B}} = \sqrt {N/M} {\mathbf{\tilde C}}$. In the improved scheme, we have ${\mathbf{G}} = \sqrt {N/M} {\mathbf{\tilde C}}{{\mathbf{\tilde C}}^{\text{H}}}$, with its off-diagonal elements
\begin{equation}
  \begin{split}
  \left| {{{\left[ {{\mathbf{\tilde C}}{{{\mathbf{\tilde C}}}^{\text{H}}}} \right]}_{p,q}}} \right| =& \left| {\sum\limits_{m = 1}^M {{e^{j\left( {{\phi _{p,m}} - {\phi _{q,m}}} \right)}}{c_{p,m}}{c_{q,m}}} } \right|\\
   \leqslant & \left| {\sum\limits_{m = 1}^M {{c_{p,m}}{c_{q,m}}} } \right| = \left| {{{\left[ {{\mathbf{C}}{{\mathbf{C}}^{\text{H}}}} \right]}_{p,q}}} \right|, p\neq q,
  \end{split}
\end{equation}
which proves that adding a random phase shift can effectively suppress side lobes.

\emph {\textbf{Pattern 4: Proposed random pattern for PBT}}

We typically desire training beam weights to have a constant modulus to ensure efficient power utilization. However, \emph{Pattern 2} and \emph{Pattern 3} do not satisfy this requirement. A simple probe pattern with constant modulus can be realized using a random phase distribution, i.e., ${b_{n,m}} = {e^{j{\varphi _{n,m}}}}/\sqrt{N}$, where $\varphi_{n,m}$ follows a uniform distribution over $[0,2\pi]$. Furthermore, by computing ${{\mathbf{D}}^{\text{H}}}{\mathbf{B}}$ and normalizing each row, we can obtain the combining pattern ${\mathbf{\tilde B}}$. The qualitative explanation for the effectiveness of the random pattern is that we can approximate ${\mathbf{B}}{{\mathbf{B}}^{\text{H}}} \approx {\mathbf{I}}$, leading to ${\mathbf{\tilde B}}{{\mathbf{B}}^{\text{H}}} \approx {{\mathbf{D}}^{\text{H}}}$, thereby achieving functionality similar to the exhaustive method.

\emph {\textbf{Pattern 5: Proposed pattern based on  P1/P2/P3 for PBT}}

\emph{Pattern 4} has the advantage of being easy to implement, but there is still a certain gap between the gain of the equivalent training beam and the DFT beam. Based on the proposed design principle, an optimization problem is formulated to find a local optimal pattern design. We still require the training beam to have constant modulus, i.e., ${b_{n,m}} = {e^{j{\varphi _{n,m}}}}$. Furthermore, to reduce the search space of optimization variables, let ${\mathbf{\tilde B}}$ be the normalized ${{\mathbf{D}}^{\text{H}}}{\mathbf{B}}$, then the optimization problem for pattern design is given as
\begin{equation}
  \begin{split}
    \mathop {\min }\limits_{\left\{ {{\varphi _{n,m}}} \right\}}  & \,{\lambda _1}f_1 + \lambda_2 f_2+{\lambda _3} f_3 \\
    {\text{s}}{\text{.t}}{\text{.}} & \;{\left\| {{{\mathbf{b}}_m}} \right\|_2} = 1,{\left\| {{{{\mathbf{\tilde b}}}_n}} \right\|_2} = 1,
  \end{split}
\end{equation}
where $f_i$ and $\lambda_i$ represents the objective function and weight of each principle, respectively. To solve this problem, automatic differentiation tools like TensorFlow can be utilized \cite{ref10}. Through gradient descent, a local optimal solution $\left\{ {\varphi _{n,m}^*} \right\}$ can be obtained. In addition, to expedite convergence, \emph{Pattern 4} can be used as the initial value.

\section{Simulation Results}
This section verifies the PBT scheme and compares it with other existing solutions. The simulation assumes that the BS is equipped with $N=256$ antenna elements, corresponding to 256 candidate DFT beams. For the geometry-based channel model, there are $P=3$ NLoS paths, where the LoS path gain ${\alpha _0} = 1$, the NLoS path gain ${\alpha _p}\sim\mathcal{C}\mathcal{N}\left( {0,0.1} \right)$, and the path angles ${\theta _p}$ obey a uniform distribution in $\left[{-1,1} \right]$. For \emph{Pattern 5}, the optimization weight is set to ${\lambda_1} = {\lambda_2} = {\lambda_3} = 1$. In the scheme based on CS, the OMP algorithm is used to achieve sparse signal recovery, and the number of pilot symbols is kept the same as PBT.

\begin{table}[h] 
  \caption{Comparison of Different Pattern Designs}
  \label{t1}
  \begin{tabular}{|c|c|c|c|c|c|}
  \hline
  \begin{tabular}[c]{@{}l@{}}Pattern\\ Design\end{tabular}                 & Overhead $M$ & \emph{P1} & \emph{P2} & \emph{P3} &  \begin{tabular}[c]{@{}l@{}}Constant\\ Modulus\end{tabular}                  \\\hline
  1                 & 256 & 0 & 0 & 0 & Yes                  \\\hline
  \multirow{2}{*}{2} & 128 & 0 & 0.0081 & 0.297 & \multirow{2}{*}{No}  \\\cline{2-5}
                    & 64  & 0  & 0.0073 & 0.250 &               \\\hline
  \multirow{2}{*}{3} & 128  &  0 & 0.0069 & 0.198  & \multirow{2}{*}{No}  \\\cline{2-5}
                    & 64  & 0  & 0.0069 & 0.248 &                    \\\hline
  \multirow{2}{*}{4} &  128 & 0.703 & 0.0078 & 0.078 & \multirow{2}{*}{Yes} \\\cline{2-5}
                    & 64 & 0.497  & 0.0078  &  0.079 &               \\    \hline
    \multirow{2}{*}{5} &  128 & $2.30\times10^{-5}$ & 0.004 & 0.039 & \multirow{2}{*}{Yes} \\\cline{2-5}
                    & 64 & $1.16\times10^{-5}$  & 0.006  &  0.051 &               \\    \hline
  \end{tabular}
\end{table}

Table \ref{t1} compares the characteristics of different pattern designs based on the proposed principles, with \emph{Pattern 1} serving as the ideal case. We focus our analysis on the other pattern designs. Compared to \emph{Pattern 2}, \emph{Pattern 3} successfully reduces energy leakage and suppresses the maximum side lobe. However, both pattern designs fail to ensure the constant modulus property of the training beams. \emph{Pattern 5}, which comprehensively considers all design principles, does not guarantee strictly equal main lobe gain but offers the best performance in terms of side lobe power leakage and maximum side lobe suppression, while also maintaining the constant modulus advantage.

\begin{figure}[t]
	\centering
	\includegraphics[width=3.2in]{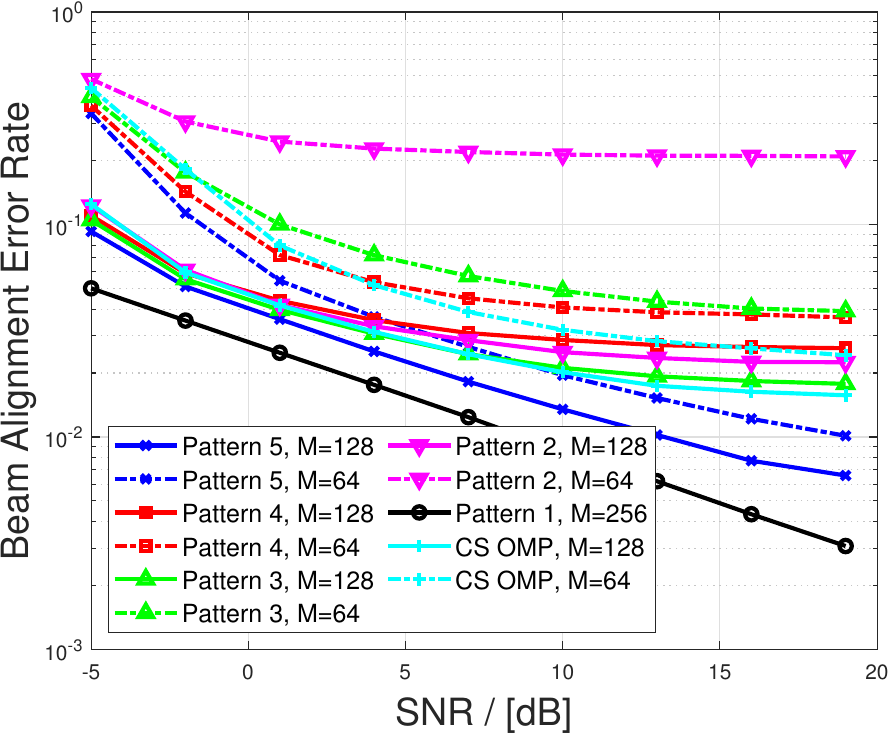}
	\caption{Beam alignment performance of different pattern designs.}
	\label{fig2}
\end{figure}

Fig. \ref{fig2} compares the beam alignment error probability $\mathbb{P}\left\{ {\hat n \ne {n^*}} \right\}$ under different pattern designs. The simulation results show a similar trend to the data in Table \ref{t1}. \emph{Pattern 3} exhibits better beam alignment performance compared to \emph{Pattern 2}, highlighting the significant impact of sidelobe suppression on performance. Moreover, \emph{Pattern 5} outperforms the existing CS scheme and other pattern designs in terms of beam alignment accuracy. Using only half the number of pilots compared to the exhaustive method, it shows about a 3 dB performance gap. This is because, under \emph{Pattern 5}, the power of the training signals is halved, which is also reflected in the beam gain shown in Fig. \ref{fig1}. Nevertheless, the proposed PBT scheme can achieve a balance between overhead and SNR conditions.

\section{Conclusion}\label{sec6}
To address the issue of excessive beam training overhead in ELAA, this paper proposes a novel PBT scheme. With only a single linear processing of the pilots at both the BS and the UE, the proposed scheme can achieve the functionality of traditional exhaustive methods while requiring only half, or even less, of the overhead. Additionally, we provide a detailed discussion on the design principles of probing patterns and combining patterns, along with specific examples. Simulation results demonstrate that the PBT scheme can achieve beam alignment with high accuracy, making it a promising alternative.

\bibliographystyle{IEEEtran}
\bibliography{mybib}

\begin{thebibliography}{10}
\providecommand{\url}[1]{#1}
\csname url@samestyle\endcsname
\providecommand{\newblock}{\relax}
\providecommand{\bibinfo}[2]{#2}
\providecommand{\BIBentrySTDinterwordspacing}{\spaceskip=0pt\relax}
\providecommand{\BIBentryALTinterwordstretchfactor}{4}
\providecommand{\BIBentryALTinterwordspacing}{\spaceskip=\fontdimen2\font plus
\BIBentryALTinterwordstretchfactor\fontdimen3\font minus
  \fontdimen4\font\relax}
\providecommand{\BIBforeignlanguage}[2]{{%
\expandafter\ifx\csname l@#1\endcsname\relax
\typeout{** WARNING: IEEEtran.bst: No hyphenation pattern has been}%
\typeout{** loaded for the language `#1'. Using the pattern for}%
\typeout{** the default language instead.}%
\else
\language=\csname l@#1\endcsname
\fi
#2}}
\providecommand{\BIBdecl}{\relax}
\BIBdecl

\bibitem{ref1}
N.~U. Saqib \emph{et~al.}, ``{THz} communications: A key enabler for future
  cellular networks,'' \emph{IEEE Access}, vol.~11, pp. 117\,474--117\,493,
  2023.

\bibitem{ref2}
Y.~Lu, Z.~Zhang, and L.~Dai, ``Hierarchical beam training for extremely
  large-scale {MIMO}: From far-field to near-field,'' \emph{IEEE Trans.
  Commun.}, vol.~72, no.~4, pp. 2247--2259, Apr. 2024.

\bibitem{ref3}
J.~{Zhang}, Y.~{Huang}, Q.~{Shi}, J.~{Wang}, and L.~{Yang}, ``Codebook design
  for beam alignment in millimeter wave communication systems,'' \emph{IEEE J.
  Sel. Areas Commun.}, vol.~65, no.~11, pp. 4980--4995, Nov. 2017.

\bibitem{ref4}
H.~Yu, P.~Guan, W.~Qu, and Y.~Zhao, ``An improved beam training scheme under
  hierarchical codebook,'' \emph{IEEE Access}, vol.~8, pp. 53\,627--53\,635,
  2020.

\bibitem{ref5}
J.~Zhang, Y.~Huang, Q.~Shi, J.~Wang, and L.~Yang, ``Codebook design for beam
  alignment in millimeter wave communication systems,'' \emph{IEEE Trans.
  Commun.}, vol.~65, no.~11, pp. 4980--4995, Nov. 2017.

\bibitem{ref6}
H.~Yan and D.~Cabric, ``Compressive initial access and beamforming training for
  millimeter-wave cellular systems,'' \emph{IEEE J. Sel. Topics in Signal
  Process.}, vol.~13, no.~5, pp. 1151--1166, Sep. 2019.

\bibitem{ref6_1}
Z.~Marzi, D.~Ramasamy, and U.~Madhow, ``Compressive channel estimation and
  tracking for large arrays in mm-wave picocells,'' \emph{IEEE J. Sel. Topics
  Signal Process.}, vol.~10, no.~3, pp. 514--527, Apr. 2016.

\bibitem{ref7}
X.~Wang, X.~Hou, L.~Chen, S.~Suyama, and T.~Asai, ``Hash-based fast beam
  alignment for 6{G} sub-terahertz {MIMO},'' in \emph{2022 27th Asia Pacific
  Conference on Communications (APCC)}, Oct. 2022, pp. 317--322.

\bibitem{ref8}
C.~You, B.~Zheng, and R.~Zhang, ``Fast beam training for {IRS}-assisted
  multiuser communications,'' \emph{IEEE Wireless Commun. Lett.}, vol.~9,
  no.~11, pp. 1845--1849, Nov. 2020.

\bibitem{ref9}
\BIBentryALTinterwordspacing
Y.~Si, H.~Yu, and Y.~Chen, ``Fast beam training and performance analysis for
  extremely large aperture array,'' arXiv:2404.18046 [eess.SP], Apr. 2024.
  [Online]. Available: \url{https://arxiv.org/abs/2404.18046}
\BIBentrySTDinterwordspacing

\bibitem{ref10}
\BIBentryALTinterwordspacing
M.~Abadi \emph{et~al.}, ``{TensorFlow}: Large-scale machine learning on
  heterogeneous systems,'' 2015, software available from tensorflow.org.
  [Online]. Available: \url{https://www.tensorflow.org/}
\BIBentrySTDinterwordspacing

\end{thebibliography}
\end{document}